\documentstyle[12pt,aaspp]{article}
\begin{document}

\title{Detection of Extended Polarized Ultraviolet Radiation From
the z$=1.82$~ Radio Galaxy 3C~256}

\author{Buell T. Jannuzi\altaffilmark{1,2}}
\affil{National Optical Astronomy Observatories, P.O. Box 26732,
Tucson, AZ 85726-6732 and Institute for Advanced Study, Princeton, N.J.}
\affil{email: jannuzi@noao.edu}

\author{Richard Elston}
\affil{National Optical Astronomy Observatories, CTIO, Casilla 603, La
Serena, Chile 1353}

\author{Gary D. Schmidt\altaffilmark{1} and Paul S. Smith\altaffilmark{1}}
\affil{Steward Observatory, University of Arizona, Tucson, AZ~~85721}

\author{H. S. Stockman}
\affil{Space Telescope Science Institute\altaffilmark{3}, Baltimore, MD~~21218}

\altaffiltext{1}{Visiting Astronomer, National Optical Astronomy
Observatories Kitt Peak National Observatory, operated by the
Association of Universities for Research in Astronomy, Inc., under
contract with the National Science Foundation.}
\altaffiltext{2}{Hubble Fellow}
\altaffiltext{3}{Operated by the Association of Universities for
Research in Astronomy, Inc., under contract with the National
Aeronautics \& Space Administration}

\begin{abstract}

We have detected spatially extended linear polarized UV emission from
the high-redshift radio galaxy 3C~256 ($z=1.82$). A spatially
integrated ($7.8''$ diameter aperture) measurement of the degree of
polarization of the $V-$band (rest frame 0.19 $\mu$m) emission yields
a value of 16.4\% ($\pm 2.2$\%) with a position angle of
$42{}\rlap{\rm .}^\circ 4$ ($\pm 3{}\rlap{\rm .}^\circ 9$), orthogonal
to the position angle on the sky of the major axis of the extended
emission.  The peak emission measured with a $3.6''$ diameter circular
aperture is 11.7\% ($\pm 1.5$\%) polarized with a position angle of
$42{}\rlap{\rm .}^\circ 4$ ($\pm 3{}\rlap{\rm .}^\circ 6$).  An image
of the polarized flux is presented, clearly displaying that the
polarized flux is extended and present over the entire extent of the
object.  While it has been suggested that the UV continuum of 3C~256
might be due to star formation (Elston 1988) or a protogalaxy
(Eisenhardt \& Dickinson 1993) based on its extremely blue spectral
energy distribution and similar morphology at UV and visible
wavelengths, we are unable to reconcile the observed high degree of
polarization with such a model.  While the detection of polarized
emission from HZRGs has been shown to be a common phenomena, 3C~256 is
only the third object for which a measurement of the extended
polarized UV emission has been presented.  These data lend additional
support to the suggestion first made by di Serego Alighieri and
collaborators that the ``alignment effect'', the tendency for the
extended UV continuum radiation and line emission from HZRGs to be
aligned with the major axis of the extended radio emission, is in
large part due to scattering of anisotropic nuclear emission.

\end{abstract}


\section{Introduction}

In this paper we present imaging linear polarimetry observations of
the high redshift radio galaxy (HZRG; $z>0.7$) 3C~256 ($z=1.819$,
Spinrad \& Djorgovski 1984).  It has been proposed as a protogalaxy
candidate because of its extremely blue spectral energy distribution
(SED) (it is the bluest 3C galaxy known, observed $R-K=3.6\pm0.3$, Le
F\'evre et al.$\,$ 1988 and Lilly, Longair, \& Allington-Smith 1985;
$R-K=3.2\pm0.3$, Elston 1988; $R-K=2.4$, Eisenhardt \& Dickinson
1993), its elongated morphology in both $R$ and $K$ band images (rest
frame 0.23 and 0.8 $\mu$m respectively), and because a 0.3 Gyr
starburst model (Eisenhardt {\it et al.$\,$} 1992) can fit the
observed SED. However, Elston (1988) has argued it is at most a star
formation episode involving less than 10\% of the galaxy's mass.

This object is also one of the best examples of the ``alignment
effect'', the strong tendency for HZRGs to have extended UV continuum
radiation and line emission roughly aligned with the major axis of
their radio emission (McCarthy {\it et al.$\,$} 1987; Chambers, Miley,
\& van Breugel 1987; optical properties of HZRGs and models for the
alignment effect have been recently reviewed by McCarthy 1993,
hereafter M93).  The proposed models for the extended UV light include
starlight from a burst of star formation induced by the interaction of
the radio jet and the interstellar medium (McCarthy {\it et al.$\,$}
1987; Chambers, Miley, \& van Breugel 1987; De Young 1989; M93).  The
leading alternative model is the scattering (by dust or electrons) of
anisotropic nuclear emission which is produced by an active galactic
nucleus (AGN).  Models using dust as the scattering medium were
proposed by di Serego Alighieri {\it et al.$\,$} (1989) and Tadhunter,
Fosbury, \& di Serego Alighieri (1989).  Fabian (1989) suggested that
scattering might also be expected by electrons in a hot halo
surrounding the radio galaxies, similar in nature to the halos around
cD galaxies.  An unavoidable consequence of these scattering models is
that the spatially extended radiation should be highly polarized with
the polarization position angle orthogonal to the major axis of the
extended emission (for the most recent review of these models and most
of the published polarimetry data see Cimatti {\it et al.$\,$} 1993).
Over the past six years spatially integrated measurements of the
polarization of HZRGs have established that significant polarized
emission is a common occurrence (cf. Tadhunter {\it et al.$\,$} 1992;
Cimatti {\it et al.$\,$} 1993), but separate measurements of the
polarization of the extended radiation (i.e. made from spatially
resolved images of the polarized emission) have previously only been
reported for two high redshift objects, 3C~368 ($z=1.132$; Scarrott,
Rolph, \& Tadhunter 1990) and 3C~265 ($z=0.811$, Jannuzi \& Elston
1992, Jannuzi 1994, Jannuzi \& Elston 1996).  A detection has also
been reported for the extended light from the moderate redshift
($z=0.567$) radio galaxy $1336+020$ (di Serego Alighieri, Cimatti, \&
Fosbury 1993). Measurements of the extended radiation from additional
HZRGs will provide a powerful discriminant between the proposed
models.

While 3C~256 exhibits the alignment effect, it is unusual because the
observed radiation is extended ($6-7''$) not only in the rest frame UV
(Spinrad \& Djorgovski 1984; Le F\'evre {\it et al.$\,$} 1988; this
paper), but also in the near IR (rest frame 0.8 $\mu$m, Eisenhardt \&
Dickinson 1993).  While this is not unprecedented, the observed IR
emission from most HZRGs is more compact than the rest frame UV
(Rigler {\it et al.$\,$} 1992).  All of this light is aligned with the
radio axis defined by the VLA radio map of van Breugel \& McCarthy
(1990), which shows the two radio lobes to be located at the ends of
the elongated $R$ band light.  Extended near IR radiation is not a
natural consequence of the dust scattering models presented by di
Serego Alighieri {\it et al.$\,$} (1993) and Cimatti {\it et al.$\,$}
(1993).  In the case of 3C~256 the UV surface brightness is very high
and since the observed near-IR emission is actually rest frame optical
light the aligned near-IR emission can simply be the extrapolation of
a flat spectrum or even blue UV component.  Together with the emission
line spectra which show strong, but narrow emission lines ($< 1500$
km/sec FWHM, Spinrad \& Djorgovski 1984; Spinrad {\it et al.$\,$}
1985), the multi-band imaging data provides a case for 3C~256 being a
galaxy observed early in its star-forming life with no evidence (other
than the radio lobe emission) of any active nuclear emission
(Eisenhardt \& Dickinson 1993).  However, such a model does not
naturally produce highly polarized extended radiation and our imaging
polarimetry observations provide a discriminant between the competing
models for the extended rest frame UV radiation from 3C~256.

\section{Observations}

Imaging linear polarimetry observations of 3C~256 were obtained on
1993 April 24 and 25 UT with the CCD imaging spectropolarimeter
described by Schmidt, Stockman, \& Smith (1992) mounted on the 4~m
Mayall Telescope of Kitt Peak National Observatory.  When used as a
linear polarimeter modulation is accomplished with a rotating
semi-achromatic $1\over2$-waveplate, chopping slowly (here every
300~s) between orthogonal states of polarization.  Simultaneous
measurement of the two beams split by a Wollaston prism ensures that
systematic variations are common to both polarization senses and
cancel in the data reduction.  The detector at the time of these
observations was a thinned, back-illuminated Texas Instruments 800 by
800 CCD with low read noise (6$e^-$) and outstanding quantum
efficiency ($>$95\% at $5500$\AA) due to UV sensitization and the
application of an antireflection coating of HFO$_2$.  At the 4m, the
imaging mode provides a plate scale of $0.259''$ per pixel and an
available field approximately 23 by 24 arcseconds. Since the position
of the object was shifted between sequences (see below), the field of
view of the combined images in Figure 1 is $32''$ by $32''$ with
decreasing signal$-$to$-$noise beyond the central $16''$ square
region.  All of our observations reported in this paper were made
using the Johnson $V$ filter.

Throughout this paper we will adopt the following notation when
referring to the Stokes parameters of the observed radiation.  The
observed linearly polarized flux images corresponding to each of the
linear Stokes vectors will be referred to as the $Q$ and $U$ images,
with the $I$ Stokes image referring to the total flux.  When we refer
to the normalized Stokes parameters (e.g. $Q/I$) we will use $q$ and
$u$. The instrumental Stokes components (i.e. measured $Q$ and $U$
prior to converting to the standard convention of positive $Q$
referring to polarized emission with a position angle of 0 degrees,
North, and positive $U$ to a position angle of 45 degrees, measured
North to East) will be indicated with a $'$ (e.g. $Q'$).

Twenty four $Q',U'$ sequences totaling four hours of integration time
were obtained in moderate seeing ($\approx1.-1.5''$ FWHM) with the
position of the object on the CCD shifted between sequences to allow
optimum flat fielding.

Observing conditions were not photometric on either night, preventing
a measurement of the $V$ band flux zero point. Calibration of the
polarization position angle utilized observations of standard CRL 2688
(Egg Nebula; Turnshek {\it et al.$\,$} 1990) obtained during the same
night.  Our measurements of positions 2 and 3 in the nebulae (see
figure in Turnshek {\it et al.$\,$}) are listed in Table~1 along with
the values from Turnshek {\it et al}.  The tabulated measurements have
been converted from the instrumental $q'$ and $u'$ Stokes vectors to
$q$ and $u$ by applying a position angle rotation of $\Delta \theta =
+42{}\rlap{\rm .}^\circ 4$ in order to bring our standard observations
into agreement with the published position angles for the calibration
source.  The high overall modulation efficiency of the instrument
($>97$ \%) is evident by comparison of the observed percent
polarizations and the published values.  The efficiency of the
instrument was also measured directly (by introducing a polarizing
prism into the beam and taking a full $Q',U'$ sequence) to be 98.8\%.

In Figure 1 we present combined images of the $V$ band Stokes $I$
(total flux), $Q$, and $U$ for 3C~256 and the eastern portion of
Wyndham's galaxy, a foreground spiral galaxy located 18$''$ to the
west of 3C~256.  Note that no emission from Wyndham's galaxy (named by
Spinrad and Djorgovski 1984) is present in either the $Q$ or $U$
images, while it is quite obvious in the $I$ (total flux) image.  As
expected, Wyndham's galaxy is not polarized. 3C~256, however, is
present in the $U$ and $I$ images while disappearing from the $Q$.
The polarized light is easier to see by eye in the images smoothed
with a Gaussian with a FWHM of $0.5''$ (smaller than the seeing disk)
and appears to be present over the entire extent of the object.

\begin{planotable}{lrllllllrr}
\tablewidth{6.75in}
\tablecaption{$V$ Band Polarimetry Measurements}
\tablehead{
\colhead{Object\tablenotemark{a}}& \colhead{Aperture\tablenotemark{b}}
&\colhead{$Q/I$}          & \colhead{$\sigma_{q}$} &
\colhead{$U/I$}          & \colhead{$\sigma_{u}$} &
\colhead{\%$P_V$}            & \colhead{$\sigma_{P}$} &
\colhead{$\theta_V$}       & \colhead{$\sigma_{\theta}$}
}

\startdata
CRL~2688 -- 2 Obs\tablenotemark{c}& 5.30 & $-0.213$ & $0.0030$
& $+0.435$ & $0.0030$ & 48.40 & 0.30 &  58.5 & 0.2     \nl
CRL~2688 -- 2 T90\tablenotemark{d}& 5.30 &&&&
& 49.74 & 0.26 & 100.7 & \nodata \nl
CRL~2688 -- 3 Obs\tablenotemark{c}& 5.30 & $-0.313$ & $0.0030$
& $+0.450$ & $0.0030$ & 54.82 & 0.30 &  62.4 & 0.2     \nl
CRL~2688 -- 3 T90\tablenotemark{d}& 5.30 &&&&& 55.19 & 0.26
& 105.0 & \nodata \nl
& & & & & & & & & \nl
3C~256 ~~~~~  Peak& 1.55 & $+0.129$ & $0.0199$ & $+0.039$
& $0.0202$ & 13.3 & 2.0 &  42.5 & 4.3     \nl
3C~256 ~~~~~  Peak& 3.63 & $+0.118$ & $0.0148$ & $+0.006$
& $0.0147$ & 11.7 & 1.5 &  42.4 & 3.6     \nl
3C~256 ~~~~~  Peak& 5.18 & $+0.129$ & $0.0162$ & $+0.008$
& $0.0163$ & 12.8 & 1.6 &  42.4 & 3.6     \nl
3C~256 ~~~~~  Peak& 7.77 & $+0.166$ & $0.0222$ & $-0.001$
& $0.0220$ & 16.4 & 2.2 &  42.4 & 3.9     \nl
3C~256 ~~~~~  Peak&10.36 & $+0.202$ & $0.0308$ & $+0.007$
& $0.0304$ & 20.0 & 3.1 &  42.4 & 4.4     \nl
&&&&&&&&&\nl
3C~256 ~~~~~  $3''$ NW& 2.00 & $+0.274$ & $0.0841$ & $+0.009$
& $0.0770$ & 26.0 & 8.4 &  42.4 & 9.3     \nl
3C~256 ~~~~~  $2''$ NW& 2.00 & $+0.172$ & $0.0373$ & $-0.004$
& $0.0355$ & 16.8 & 3.7 &  42.3 & 6.4     \nl
3C~256 ~~~~~  $1''$ NW& 2.00 & $+0.135$ & $0.0205$ & $+0.004$
& $0.0198$ & 13.4 & 2.1 &  42.4 & 4.4     \nl
3C~256 ~~~~~  Peak& 2.00 & $+0.115$ & $0.0160$ & $+0.034$
& $0.0152$ & 11.9 & 1.6 &  42.5 & 3.9     \nl
3C~256 ~~~~~  $1''$ SE& 2.00 & $+0.107$ & $0.0203$ & $+0.016$
& $0.0190$ & 10.6 & 2.0 &  42.5 & 5.5     \nl
3C~256 ~~~~~  $2''$ SE& 2.00 & $+0.112$ & $0.0409$ & $-0.040$
& $0.0392$ & 11.2 & 4.1 &  42.2 &10.5     \nl
\tablenotetext{a}{The calibration source CRL~2688 was measured at two
positions (identified as positions ``2'' and ``3'' in
Turnshek {\it et al.$\,$}
1990 (T90). Both our observed instrumental values (indicated with
``Obs'') and the Turnshek {\it et al.$\,$} values (indicated with T90) are
listed.  For 3C~256, ``Peak'' indicates the photometry aperture was
centered on the peak of the total $V$ band emission. Positional
offsets in arcseconds and relative to the position of the peak
emission are listed for some measurements}
\tablenotetext{b}{Aperture diameter in arcseconds}
\tablenotetext{c}{Uncalibrated instrumental unit measurements of
CRL~2688 used to
calibrate the position angle offset.}
\tablenotetext{d}{The degree of polarization and position
angle of the polarized radiation observed by Turnshek {\it et al.$\,$}
1990 for positions ``2'' and ``3'' of CRL~2688.}
\end{planotable}

In Table~1 we have tabulated measurements of the percent polarization
made from the $Q$, $U$, and $I$ images for various positions in 3C~256
and with a range of aperture sizes.  All the tabulated measurements
were made on the unsmoothed images from the combined data set (both
nights) using the IRAF APPHOT task to measure the $Q$, $U$, and $I$
fluxes.  Measurements were made not only on the combined data, but
also on four subsets of data, selected to be half of each nights'
data. All four measurements of the normalized Stokes vectors are
consistent within the one sigma uncertainties of the measurements made
on the combined data set.  The uncertainties of all our measured $Q$~
and $U$~ fluxes listed in Table~1 are dominated by the uncertainty in
the measured background due to the sky and flat fielding residuals.
For this reason we determined the variance of the background
empirically for each measurement, using two methods which proved in
all cases to be consistent.  The first method used an annulus around
the aperture always chosen to be large enough to include ``sky''
pixels, but avoid light from the object.  The second method used was
the random placement over the image of apertures identical in size to
the ``object'' aperture.  The variance of these measurements, after
rejection of apertures including 3C~256 and Wyndham's galaxy, were
consistent with the variance determined from the measurements using
the surrounding annulus.  The first set of tabulated measurements all
have the aperture centered on the peak of the radiation in the map of
the total flux. Note that the percent polarization remains high or
increases as the aperture size increases and includes more extended
radiation.  The second set of tabulated measurements were all made
with the same aperture size ($2''$ diameter), but with the center of
the aperture moved in $1''$ increments along the major axis of the
extended radiation.  While the spatial offsets of the apertures for
adjacent tabulated measurements are smaller than the diameter of the
aperture, the $3''$ NW, $1''$ NW, and $1''$ SE measurements compose a
group of three spatially independent measurements of the percent
polarization over a linear extent of $6''$.  Note that the measured
peak of the polarized flux is located $0.23''$ W and $0.4''$ N of the
peak in the total flux map, along the major axis of the optical and
radio emission.

\makeatletter
\def\jnl@aj{AJ}
\ifx\revtex@jnl\jnl@aj\let\tablebreak=\nl\fi
\makeatother

Limited spatial resolution and unpolarized narrow line emission
(strong CIII] (1909\AA) falls in the middle of our passband) can
dilute the measurement of the polarized continuum radiation.  Assuming
that the extended line emission from 3C~256 is unpolarized means the
intrinsic polarization of the observed UV continuum is larger than the
values listed in Table~1. However, the observed equivalent width of
the CIII] emission is only 79 \AA~ (28\AA~ in the rest frame, Spinrad
{\it et al.$\,$} 1985), so this would only be a correction on the
order of 5\% (for example, an increase of the degree of polarization
from 16\% to 17\%).

The major axis of the extended radiation has a position angle on the
sky of approximately $138^\circ$ ($132^\circ$) (measured respectively
by Spinrad \& Djorgovski 1984 and the current study).  All of the
observed polarized light has a position angle orthogonal to this axis
(difference $\approx 95^\circ$ to $91^\circ$).

\section{The Polarizing Mechanism}

The high degree of polarization that we observe for the $V$-- band
light from 3C~256 makes it very unlikely that dichroic extinction by
aligned dust grains is responsible for producing the observed
polarization. Synchrotron emission could produce such high degrees of
polarization.  The lifetime for electrons emitting at UV wavelengths
is short.  To produce all of the polarized light along the entire
extent of 3C~256 (a linear distance of 20 to 40 kpc depending on the
choice of cosmology) and assuming a central AGN as the original source
for the charged particles, would require a mechanism to reaccelerate
the particles over this entire region.  In addition, because of the
lack of any variation in the position angle, the magnetic field would
have to be ordered in a similar geometry along the entire extent of
the object, unlike what is observed in highly polarized
optical-synchrotron jets (e.g. M87, 3C~273; although spatial dilution
might make it more difficult to detect such changes if they were
present).

This leaves scattering, either by electrons or dust, as the probable
mechanism for producing the polarized flux.  Unfortunately with only a
single observed passband we can not determine the wavelength
dependence of the polarized flux, which would would be a powerful
discriminant between dust (whose cross section decreases with wavelength)
and electrons (a grey scatterer) as the scattering
medium. Circumstantial evidence would favor dust scattering.  The
unusual blue color of the radiation has already been noted (Eisenhardt
\& Dickinson 1993). In addition, Le F\'evre {\it et al.$\,$} (1988)
obtained high spatial resolution ($0.7''$) $R$ and $I$ band images
which resolved the 3C~256 into three components which they labeled
``a'' (central), ``b'' (SE component), and ``c'' (NW component).  In
our image the ``a'' and ``b'' components are certainly merged and have
become what we call in Table~1 the ``peak'' of the observed
radiation.Le F\'evre {\it et al.$\,$} claim that the object gets bluer
as one moves toward the ends of the elongated radiation,
i.e. component ``a'' is redder than ``b'' or ``c''.  In our data it is
not possible to determine if the degree of polarization also increases
away from the peak radiation.  Such behavior would be the expected
consequence of dust scattered light producing an increasing fraction
of the observed light with increasing distance from the center of the
galaxy where a compact stellar population might be diluting the
percent polarization.  However, we caution that the significance of
the color gradient is unclear. Eisenhardt \& Dickinson (1993) and
Eisenhardt {\it et al.$\,$} (1992) report no evidence for any color
gradient and the published uncertainties in the Le F\'evre {\it et
al.$\,$} paper are large.  More accurate, high spatial resolution maps
of the color of 3C~256 would help determine the nature of the
radiation.

While the measured broad band colors of the total and polarized flux
do not yet allow discrimination between dust or electron scattering,
we have attempted to used the measured amount and spatial distribution
of the polarized flux as a diagnostic of the scattering mechanism.  At
the rest wavelengths investigated by our observations, the cross
section for scattering per unit mass is three magnitudes larger for
dust (assuming Galactic, LMC, or SMC distributions of dust grain
sizes) than for electrons. If the total amount of mass necessary to
produce the scattered radiation by electron scattering were absurdly
large then the case for identifying dust as the scatters would be
strengthened.  First we considered the most favorable case for
electron scattering. We chose a favorable cosmology (so that the total
size of 3C~256 is as small as possible ($H_\circ = 75$, $q_\circ =
0.5$), bringing the scatters as close to the central source as
possible), a luminous central AGN (one bright enough that if viewed
directly it would have a $V$ magnitude of 16.5), and a scattering
geometry which scatters the maximum amount of light into our line of
site and produces 100\% polarized scattered light.  We then derived the
electron mass in each image pixel from its polarized flux and distance
from the central source. Summing the mass in all the pixels within the
29.6 magnitude per square arcsecond isophot (in the total flux image,
with the zeropoint set by the photometry of Eisenhardt and Dickinson
1992) gives a lower limit to the total mass of more than $1.1 \times
10^9 M_\odot$.  This is a firm lower limit.  A variety of effects
could result in an increase in the mass estimate.  If the intrinsic
polarization of the scattered flux were only 40\% then the needed mass
would increase. Spatial dilution almost certainly has resulted in a
measured value for the polarized flux which is lower than the
intrinsic value.  If we assume that the scatterers are spherically
distributed and that the resulting distribution of the polarized
emission is because the AGN only illuminates a cone of half angle
$30^\circ$, then we are ``seeing'' only about 1/7 of the scatters and
the total mass increases to more than $2 \times 10^{10} M_\odot$.
However, such masses are still not so large that we can rule out
electron scattering as being implausible.

There is some circumstantial evidence that supports dust scattering
over electron scattering.  For the only two objects for which
extensive multi-band polarimetry is available 3C~368 (Cimatti {\it et
al.$\,$} 1993) and 3C~265 (Jannuzi \& Elston 1991; Jannuzi 1994;
Jannuzi \& Elston 1996), the polarized flux is blue and dust
scattering is the favored mechanism.  Nevertheless, the exact
mechanism responsible in the case of 3C~256 remains uncertain and
additional broad-band polarimetry and/or spectropolarimetry is
required.

\section{The Alignment Effect in 3C~256 and Summary}

Our detection of highly polarized extended radiation from 3C~256 is
not consistent with simple starburst models for the production of the
extended and radio-aligned UV light. Like all of the HZRGs with
detected polarized light (cf. Cimatti {\it et al.$\,$} 1993), the
position angle of the polarized flux from 3C~256 is orthogonal to the
major axis of the UV radiation.  A simple source geometry that could
produce such a situation is a compact nuclear source illuminating
scatters located along the region containing the extended light.
However, if the extended UV light is produced by the scattering of a
nuclear source, what is the nature of that nuclear emission?  Given
the high polarization of the UV continuum radiation from 3C~256, more
than 30\% of the extended radiation must be scattered light from a
anisotropic source which is not directly visible along our line of
sight. If the anisotropic source is intrinsically similar to a radio
loud quasar and dust is the scattering medium, then high
signal-to-noise ratio spectroscopy should reveal a broad line region
(even if not observable directly the dust should scatter some of the
broad line region emission into our line of sight).  If hot electrons
are the scatters, the broad line region emission might be broadened
until it is undetectable. However, in that case the slope of the
spectrum of polarized flux would be a direct reflection of the
spectral energy distribution of the nuclear light and 3C~256 should
have the colors of a typical quasar.  Additional broad-band and/or
spectropolarimetry is required to unambiguously determine the
scattering mechanism and intrinsic nature of the source producing the
emission incident on the scatters.

\acknowledgments

We thank the staff of KPNO, especially Skip Andre, for providing the
technical assistance necessary to ensure smooth operation of the
spectropolarimeter on the Mayall 4~m telescope. We acknowledge
stimulating and instructive conversations with Sperello di Serego
Alighieri, Mark Dickinson, Peter Eisenhardt, \& Pat McCarthy.  This
research was partially supported by NASA grant NAG~5--1630 and NSF
grant AST~91--14087.  B.T.J. acknowledges support from NASA through
grant number HF--1045.01--93A from the Space Telescope Science
Institute, which is operated by the Association of Universities for
Research in Astronomy, Incorporated, under NASA contract NAS5--26555.

\bigskip\bigskip

\section*{Figure Captions}

\noindent
Fig. 1.\ Images of the Stokes $I$ (a), $Q$ (b), and $U$ (c) fluxes are
presented. For all images North is up, East is to the left.  The plate
scale is $0.259''$ per pixel.  The field of view for the top three
frames is $29''$ by $29''$.  a.) An image of the total $V$-band flux
from 3C~256, centered in the frame, and Wyndham's galaxy to the west.
b.) The $Q$ flux image corresponding to a. c.) The $U$ flux image
corresponding to a. d-f.)are magnified versions of (a-c) to better
display 3C~256. The field of view is $15.5''$ by $15.5''$ arcseconds.
The bottom row of images (g-h)are the same as (d-f), but convolved
with a Gaussian with a FWHM of $0.5''$. The images are negatives and
observed radiation in both the total and polarized flux images appears
black.

\end{document}